\documentstyle[11pt]{article}
\def\be{\begin{equation}}
\def\ee{\end{equation}}

\begin{document}

\begin{flushright}
CERN-TH/99-278\\
RIKEN-BNL preprint\\
UT-Komaba preprint
\end{flushright}
\begin{center}
{\Large\bf{Scalar Glueball--Quarkonium Mixing and the Structure of the QCD
Vacuum}}
\vskip1cm
{\bf John Ellis}$^{a}$,
{\bf Hirotsugu Fujii}$^{b}$ and {\bf Dmitri Kharzeev}$^{c}$
\vskip0.3cm
{\it{
$^{a}$ Theory Division, 

CERN, 

Geneva 23, CH--1211, Switzerland

\vskip0.2cm
$^{b}$ Institute of Physics, University of Tokyo at Komaba, 

3-8-1 Komaba, Tokyo 153-8902, Japan

\vskip0.2cm
$^{c}$ RIKEN--BNL Research Center, 

Brookhaven National Laboratory, 

Upton NY 11973, USA
}}
\end{center}
\vskip0.5cm
\begin{abstract}
We use Ward identities of broken scale invariance to infer   
the amount of scalar glueball--$\bar{q}q$ meson mixing from the
ratio of quark and gluon 
condensates in the QCD vacuum. Assuming dominance by a single scalar
state, as suggested by a phase-shift analysis,
we find a mixing angle $\gamma \sim 36^{\circ}$, corresponding to
near-maximal mixing of the glueball and $\bar{s}s$ components.
\end{abstract}

\vskip0.5cm

Many scalar mesons are predicted in non-perturbative QCD, 
including $\bar q q$ bound
states, glueballs~\cite{Teper}, $\bar q q \bar q q$ molecules, 
radial excitations, and hybrids.
Experimentally, there have also been many 
reported sightings of scalar states,
including the $\sigma(400-700)$, $f_0(980)$,
$f_0(1300)$, $f_0(1500)$, $f_0(1700), \dots$~\cite{PDG}. The
theoretical identifications of these states are still largely open, in
particular the identification of the lightest scalar glueball.  This quest is
complicated by the expectation that the various different scalar states
could
mix with each other~\cite{Lee}, sharing out any characteristic glueball
signatures and
polluting even the strongest candidates~\cite{Kirk} with, e.g., $\bar q q$
features~\cite{Ochs}.

One approach to the scalar-meson problem offered by non-perturbative field
theory is based on the consideration of Green functions of composite operators
such as $\bar q(x)q(x)$ or $G^a_{\mu\nu}(x)G^{a\mu\nu}(x) \equiv G^2$,
constrained by
the
(approximate and/or anomalous) Ward identities of non-perturbative symmetries of
QCD.  The highly successful prototype for this approach has been chiral
symmetry, and it has also often been applied to broken
scale invariance \cite{JE,JE1,sch,EL}, with some success.

Unlike chiral symmetry, where the Ward identities are dominated by low-mass
pseudoscalar mesons, there is no guarantee that the Ward identities of broken
scale invariance should be dominated by any single scalar-meson state.  Under
these circumstances, the best formulation of the approach may be
phenomenological, setting up a number of sum rules~\cite{NSVZ} based
on the Ward identities of broken scale
invariance~\cite{scale}, and substituting experimental data into
them, with the aim of
exploring empirically which collection of states may saturate them.

We have recently implemented such a programme for Green functions involving the
trace of the anomalous pure QCD energy-momentum tensor $\theta {(x)}
\equiv
\frac{\beta}{4\alpha_s}G^a_{\mu\nu}(x)G^{a\mu\nu}(x)$~\cite{scale},
evaluating the sum rules
with data on $\pi\pi$ and $\bar K K$ phase shifts and phenomenological
parametrizations of observed scalar mesons~\cite{FK,EFK}.  We have found
empirically that the sum rules
are probably dominated by the $f_0(980)$ state, with contributions from the
lighter $\sigma$ and heavier $f_0$ mesons each contributing around the 10\%
level.

Does this mean that one should identify the $f_0(980)$ as the lightest scalar
glueball?  Certainly not until one has analyzed the pattern of 
mixing with $\bar
q q$ states, a complicated issue which we broach in this paper.
\vskip0.3cm

We study simple Ward identities for the two-point Green functions of
$\theta$ and
$\bar q q$ operators.  Assuming that the latter are dominated by the
$f_0(980)$, as in
the $\theta\theta$ case, we find that the mixing of this ``glueball" with an
$\bar q q$ meson must be large. For a nominal choice of vacuum parameters: 
$\langle 0|\bar s s|0\rangle = 0.8 \langle 0 |\bar q q| 0 \rangle$,
where $\bar q q$ represents either $\bar u u$ or $\bar d d$, $\langle
0
|\bar q q|0 \rangle = 0.016$~GeV$^3$ and $\langle 0
\left|\frac{\beta}{4\alpha_s}G^2 \right|0\rangle = 0.013$~GeV$^4$, we find
near-maximal mixing between glueball and $\bar s s$ components: $\gamma
\sim 36^{\circ}$, so that the $f_0(980)$ contains an
almost
equal mixture of glueball and $\bar s s$ states. 
\vskip0.3cm

We start by considering the low--energy theorem \cite{NSVZ}
\be
\lim_{q\to 0} i \int dx\ e^{iqx} \langle 0| T\left\{ 
{\beta(\alpha_s) \over 4 \alpha_s} G^2(x) ,\ {\cal{O}}(0)
\right\} |0 \rangle = 
(-d) \langle {\cal{O}} \rangle + O(m_q), \label{off}
\ee
where $d$ is the canonical dimension of the operator $\cal{O}$, 
$\beta(\alpha_s)$ is the Gell-Mann--Low function:
$\beta(\alpha_s) \equiv - b \alpha_s^2 / 2 \pi + O(\alpha_s^3), \ b = 
(11 N_c - 2 N_f)/3$,  
and $O(m_q)$ stands for the terms linear in light quark masses. 
Here and subsequently, we work only with renormalization-group
invariant quantities.

Next we use a spectral representation for the theorem (\ref{off}),
assuming that $\cal{O}$ is a scalar Hermitian operator:
\begin{eqnarray}
&&\langle 0 | T \{
{\beta(\alpha_s) \over 4 \alpha_s} G^2(x) ,\ {\cal{O}}(0) \} |0 \rangle
\nonumber \\
&=&
\int d^4k(2\pi)^{-3}\frac{1}{\pi}A(k) e^{-ikx}\theta(x^0)+
\int d^4k(2\pi)^{-3}\frac{1}{\pi}B(k) e^{ikx}\theta(-x^0),
\end{eqnarray}
where
\begin{eqnarray}
\frac{1}{\pi}A(k)&\equiv&
(2\pi)^3\sum_n \delta^4(p_n-k)
\langle 0| \frac{\beta}{4\pi}G^2(0)|n\rangle \langle n| {\cal O}(0) |0\rangle,
\\
\frac{1}{\pi}B(k)&\equiv&
(2\pi)^3\sum_n \delta^4(p_n-k)
\langle 0| {\cal O}(0) |n\rangle \langle n| \frac{\beta}{4\pi}G^2(0)|0\rangle.
\end{eqnarray}
Specializing to the relevant case where
${\cal O}(x)$ is scalar, 
it is clear that $A(k)$ and $B(k)$ are the scalar functions of the form,
\begin{equation}
A(k)\equiv A(k^2) \theta(k^0), \quad
B(k)\equiv B(k^2) \theta(k^0),
\end{equation}
where the support of the spectral condition requires the factors
$\theta(k^0)$.
Assuming also that ${\cal O}$ is CP even, as in the cases
of $\theta$ and
scalar $\bar q q$ densities, we may use time-reversal invariance
to infer that
\be
A(k^2)=B(k^2)\equiv  \rho_{\cal O}(k^2).
\ee
Similarly, it can be shown that $A(k)$ is real.

Inserting the resulting spectral representation
\begin{eqnarray}
& &\langle 0 | T \{ 
{\beta(\alpha_s) \over 4 \alpha_s} G^2(x) ,\ {\cal{O}}(0)
\} |0 \rangle
\nonumber \\
&=&
\frac{1}{\pi} \int \frac{d^4 k}{(2\pi)^3} \rho_{\cal O}(k^2)\theta(k^0)
\{ e^{-ikx}\theta(x^0)+e^{ikx}\theta(-x^0)\}
\nonumber \\
&=&
\frac{1}{\pi} \int_0^\infty dm^2 \rho_{\cal O}(m^2)
\int \frac{d^4 k}{(2\pi)^3} \delta(k^2-m^2)\theta(k^0)
\{ e^{-ikx}\theta(x^0)+e^{ikx}\theta(-x^0)\}
\nonumber \\
&=&
\frac{1}{\pi} \int_0^\infty dm^2 \rho_{\cal O}(m^2)
\Delta_F(x;m^2)
\end{eqnarray}
into the theorem (\ref{off}), we find the simple relation
\begin{equation}
\frac{1}{\pi} \int_0^\infty \frac{dm^2}{m^2} \rho_{\cal O}(m^2)
=(-d)\langle {\cal O} \rangle +O(m_q),
\end{equation}
whose physical properties we now discuss in more detail.
\vskip0.3cm

In general, the intermediate states created by the operator
${\cal O}$ may include
multi-particle states as well as single-particle
states. If one approximates the intermediate states
by a sum over single-particle states, one finds
that the matrix elements here are scalar and can depend only on 
$k^2=m_\sigma^2$ in this case.   
Then the three-momentum integral is trivially done:
\begin{eqnarray}
\frac{1}{\pi}\rho_{\cal O}(k^2) \theta(k^0)
&=& \sum_{\sigma}\delta(m_\sigma^2-k^2)\theta(k^0)
\langle 0| {\beta(\alpha_s) \over 4 \alpha_s} G^2 | k;\sigma \rangle
 \langle k;\sigma | {\cal O} | 0\rangle ,
\end{eqnarray}
where the index $\sigma$ specifies the species of scalar meson,
and $|k;\sigma\rangle$ stands for a state of momentum $k$.
The theorem (\ref{off}) therefore becomes
\be
\sum_\sigma \frac{1}{m_\sigma^2}
\langle 0| {\beta(\alpha_s) \over 4 \alpha_s} G^2 | k;\sigma \rangle
 \langle k;\sigma | {\cal O} | 0\rangle 
= (-d)\langle {\cal O}\rangle ,
\ee
which may further be simplified to
\be
\frac{1}{m_\sigma^2}
\langle 0| {\beta(\alpha_s) \over 4 \alpha_s} G^2 | k \rangle
 \langle k | {\cal O} | 0\rangle 
= (-d)\langle {\cal O}\rangle 
\ee
in the case of the single sharp resonance.

We now show how the relation (\ref{off}) can be used to fix the mixing 
between the scalar quark--antiquark and glueball states.    
We first choose ${\cal O }(x) = \sum_i m_i \bar{q_i}q_i(x)$, 
for which the 
spectral representation (\ref{off}) becomes~\footnote{We recall that the
canonical quantum operator
dimension of $\bar q q$ is 3, whereas the renormalization-group invariant 
combination $m\bar q q$ has mass scaling dimension 4~\cite{NSVZ}.
We ignore
operator mixing between $G^2$ and $\bar q q$, assuming a
mass-independent renormalization scheme such as $\overline {MS}$.}:
\be 
{1 \over \pi} \int ds\ {\tilde{\rho}(s) \over s} \simeq -3 \ 
\langle \sum_i m_i \bar{q_i}q_i \rangle, \label{dispoff}
\ee
where
\be
\frac{1}{\pi}\tilde{\rho}(s) 
=(2\pi)^3 \sum_{n}\delta^4(p_n-k)
\langle 0| {\beta(\alpha_s) \over 4 \alpha_s} G^2 | n \rangle
\langle n | \sum_i m_i \bar{q_i}q_i | 0 \rangle, 
\label{offdiag}
\ee
These relations demonstrate that the spontaneous breaking of chiral 
symmetry, reflected in a non-zero value of the quark condensate
$\langle 0 |\bar{q}q | 0\rangle$
necessarily implies mixing between the ``glueball" and ``quarkonium" 
components of physical scalar resonances $\sigma$.
In the case of the single sharp resonance $\sigma$, 
the theorem (\ref{offdiag}) implies that
\be
\frac{1}{m_\sigma^2}
\langle 0| {\beta(\alpha_s) \over 4 \alpha_s} G^2 | k\rangle
 \langle k |  \sum_i m_i \bar{q_i}q_i | 0\rangle 
=-3 \langle  \sum_i m_i \bar{q_i}q_i \rangle.
\label{qqbarsr}
\ee
Choosing instead ${\cal O}(x) = {\beta(\alpha_s) \over 4 \alpha_s} G^2(x)$
leads to another sum rule~\cite{NSVZ}:
\be
{1 \over \pi} \int ds\ {\rho (s) \over s} \simeq (-4) \
\langle {\beta(\alpha_s) \over 4 \alpha_s} G^2  \rangle, \label{diag1}
\ee
where
\be
\frac{1}{\pi}\rho (k^2)\theta(k^0) = 
(2\pi)^3 \sum_{n}\delta^4(p_n-k)
| \langle n |  {\beta(\alpha_s) \over 4 \alpha_s} G^2 | 0 \rangle |^2
, 
\label{dispdiag}
\ee
which we now analyze together with (\ref{qqbarsr}).

Combining the relations (\ref{diag1}) and (\ref{dispoff}) 
and discarding possible multi-particle intermediate states,
we find the following general
relation:
\be
{\sum_{\sigma} 
\langle 0 | {\beta(\alpha_s) \over 4 \alpha_s} G^2 | \sigma \rangle
\langle \sigma | {\beta(\alpha_s) \over 4 \alpha_s} G^2 | 0 \rangle 
/ m_\sigma^2
\over 
{\sum_{\sigma} 
\langle 0 | {\beta(\alpha_s) \over 4 \alpha_s} G^2 | \sigma \rangle
\langle \sigma |  \sum_i m_i \bar{q_i}q_i | 0 \rangle}
/ m_\sigma^2
}
 = {4 \over 3} 
{{\langle {\beta(\alpha_s) \over {4 \alpha_s}} G^2  
\rangle} 
\over {\langle \sum_i m_i \bar{q_i}q_i \rangle}}. \label{intrel1}
\ee
In the case of two flavors and neglecting the breaking of isospin
symmetry, (\ref{intrel1}) leads to
\be
{\sum_{\sigma} 
\langle 0 | {\beta(\alpha_s) \over 4 \alpha_s} G^2 | \sigma \rangle
\langle \sigma | {\beta(\alpha_s) \over 4 \alpha_s} G^2 | 0 \rangle 
/ m_\sigma^2
\over 
{\sum_{\sigma} 
\langle 0 | {\beta(\alpha_s) \over 4 \alpha_s} G^2 | \sigma \rangle
\langle \sigma | {\bar u} u + {\bar d} d | 0 \rangle}
/ m_\sigma^2
}
 = {4 \over 3} 
{{\langle {\beta(\alpha_s) \over {4 \alpha_s}} G^2  
\rangle} 
\over {\langle {\bar u} u + {\bar d} d \rangle}}. \label{intrel}
\ee
In the case of three flavors, we know that 
$m_s \gg m_u, m_d$, and the ${\bar s} s$ condensate 
is comparable to that of ${\bar u} u, {\bar d} d$~\cite{Ioffe}.
Hence the denominator of the right-hand side of (\ref{intrel1}) must be
dominated by the strange-quark contribution. If we further assume
that the scalar strange contents of scalar mesons are of 
the same order of magnitude as those involving $u$ and $d$
quarks~\footnote{A naive analysis of the $\pi$-nucleon $\sigma$ term
indicates that
$\langle N|\bar s s|N\rangle$ is not much smaller than
$\langle N |\bar q q| N \rangle$, but there is no comparable
information concerning scalar mesons.},
then the
denominator on the left-hand side of (\ref{intrel1}) will 
also be dominated by the strange-quark 
contributions, and (\ref{intrel}) can be re-written as 
\be
{\sum_{\sigma} 
\langle 0 | {\beta(\alpha_s) \over 4 \alpha_s} G^2 | \sigma \rangle
\langle \sigma | {\beta(\alpha_s) \over 4 \alpha_s} G^2 | 0 \rangle 
/ m_\sigma^2
\over 
{\sum_{\sigma} 
\langle 0 | {\beta(\alpha_s) \over 4 \alpha_s} G^2 | \sigma \rangle
\langle \sigma |  \bar s s | 0 \rangle}
/ m_\sigma^2
}
 = {4 \over 3} 
{{\langle {\beta(\alpha_s) \over {4 \alpha_s}} G^2  
\rangle} 
\over {\langle \bar s s \rangle}}. \label{intrels}
\ee
The relations (\ref{intrel1}), (\ref{intrel}), (\ref{intrels}) demonstrate 
that the ratio of the glueball and scalar $\bar{q}q$ components in the 
physical scalar resonances is determined by the ratio of the gluon and 
quark condensates in the vacuum, which is the key
theoretical foundation of this paper.
\vskip0.3cm

We now turn to the phenomenological analysis of the above sum rules.
Here, the key observation coming from the evaluation of
the sum rule (\ref{diag1}) using 
the available experimental data on $\pi\pi$ and $\bar{K}K$ phase shifts 
is that the dominant contribution to the spectral integral 
in (\ref{dispoff}) is due to the $f_0(980)$
resonance~\cite{FK,EFK}. 
It therefore makes sense to consider approximate relations which follow 
from (\ref{intrel1}) in the case when both sum rules (\ref{dispdiag}) 
and (\ref{dispoff}) are saturated by a single resonance, 
namely the $f_0(980)$~\footnote{Similar conclusions on large
glueball-quarkonium mixing would hold if any other single meson
state dominated the sum rules.}.
Since the $f_0(980)$, with a width of $\Gamma = 40 \div
100$~MeV~\cite{PDG},
is relatively narrow compared with its mass
and its separation from other scalar mesons,
it is a reasonable approximation to
approximate its spectral shape by a delta function.
The sum rule (\ref{diag1}) then can be written as 
\be
 {1 \over m_{f_0}^2}\  
| \langle f_0 | {\beta(\alpha_s) \over 4 \alpha_s} G^2 | 0 \rangle |^2
 \simeq -4 \
\langle {\beta(\alpha_s) \over 4 \alpha_s} G^2  \rangle. \label{fdiag}
\ee
Since the quantity on the right-hand side, namely the gluon condensate,
has been estimated from
QCD sum rule analyses~\cite{SVZ}, (\ref{fdiag}) fixes the coupling 
of the $f_0(980)$ resonance to the scalar glueball current. 

Because $m_s \gg m_u, m_d$, 
it is reasonable to neglect the
${\bar u} u + {\bar d} d$ contribution to the left-hand side
of (\ref{intrel1}), and use the three--flavor relation (\ref{intrels}) 
to determine the coupling of this resonance to 
the scalar quark--antiquark current 
$\bar{s}s(x)$:~\footnote{We assume that the matrix elements in 
(\ref{offdiag}) and (\ref{dispdiag}) are real, which must
be the case for non-degenerate single-particle states.}
\be 
{\langle f_0 | {\beta(\alpha_s) \over 4 \alpha_s} G^2 
| 0 \rangle \over {\langle f_0 |  \bar{s}s 
| 0 \rangle}} \simeq {4 \over 3}\ {{\langle {\beta(\alpha_s) 
\over {4 \alpha_s}} G^2  
\rangle} 
\over {\langle \bar{s}s \rangle}}, \label{intrels1}
\ee 
which we now use to quantify glueball-quarkonium mixing in
this state.
\vskip0.3cm

We assume that the $f_0$ wave function is a
superposition of glueball $| G \rangle$ and quark--antiquark 
$| \bar{Q}Q \rangle$ components~\footnote{Analogous
arguments leading to large mixing could also be made if the
$f_0(980)$ state turned out to be ${\bar q} q {\bar q} q$
molecule, as sometimes argued.}:
\be
| f_0 \rangle = \cos\ \gamma \ | G \rangle + \sin\ \gamma \ | \bar{Q}Q
\rangle. 
\label{mix}
\ee
We next assume that
the quark-antiquark component $| \bar{Q}Q \rangle$
is mainly $| \bar{S}S \rangle$: if there is a substantial
$| \bar{U}U + \bar{D}D \rangle$ component, this would (barring
a cancellation) tend to increase the estimate of the mixing angle
$\gamma$ given below. 
We 
further assume that
the glueball component $| G \rangle$ has the dominant coupling to
the scalar 
gluon current, ${\beta(\alpha_s) \over {4 \alpha_s}} G^2$, and that the
quark--antiquark component $| \bar{S}S \rangle$ has the dominant
coupling to the $\bar{s}s$ current. If this were not the
case, there would be additional mechanisms for large glueball-quarkonium
mixing that would be difficult to quantify.
Extracting a simple dimensional factor, these approximations
imply that
\be 
{\langle f_0 | {\beta(\alpha_s) \over 4 \alpha_s} G^2 
| 0 \rangle \over {\langle f_0 |  \bar{s}s 
| 0 \rangle}} = m_{f_0} \ {\rm cotan} \, \gamma,
\ee
and (\ref{intrels1}) can now be used to determine the magnitude of the mixing 
angle in (\ref{mix}):
\be
\tan\ \gamma = m_{f_0}\ {3 \over 4}\ {\langle \bar{s}s \rangle \over 
{\langle {\beta(\alpha_s) \over {4 \alpha_s}} G^2  
\rangle}}. \label{result}
\ee
Numerically, using $\langle \bar{s}s \rangle = (0.8 \pm 0.1) 
\langle \bar{q}q \rangle$ \cite{Ioffe}, 
$\langle \bar{q}q \rangle \simeq 0.016\ \rm{GeV}^3$
and $\langle {\beta(\alpha_s) \over {4 \alpha_s}} G^2  
\rangle \simeq 0.013\ \rm{GeV}^4$, 
we estimate on the basis of (\ref{result}) that
$\gamma \simeq 36^{\circ}$, i.e., the mixing between the quark-antiquark 
and the glueball components is strong, even close to maximal.
\vskip0.3cm

To summarize, we have found that the mixing between the glueball 
and $\bar{q}q$ components in the lightest scalar state dominating
sum rules for $\theta = {\beta(\alpha_s) \over 4 \alpha_s} G^2$, i.e.,
the best candidate for the lightest scalar ``glueball",
is required 
by the the ratio of the quark and gluon condensates to be very strong.
We have evaluated this mixing for the $f_0(980)$ state that has been found
empirically in a phase-shift analysis~\cite{FK,EFK} to dominate the sum
rule for $\theta \theta$, and found it to be near-maximal.
We note that this conclusion would only be strengthened if the sum
rules were to be saturated by a state heavier than the $f_0(980)$.

Our analysis has, admittedly, been rather crude. However, we feel
that it has demonstrated the power of sum rules derived from
broken scale invariance to contribute to the debate concerning
the identification of the lightest scalar glueball, indicating, in
particular, that its mixing with a quark-antiquark state may not be
neglected.

\begin{center}
{\bf Acknowledgements}
\end{center}

We take this opportunity to recall with respect our friend
and colleague Jozef Lanik, and dedicate this paper to his memory. 
We thank M.S. Chanowitz for a useful discussion.
H.F. and D.K. thank RIKEN, Brookhaven National Laboratory and 
the U.S. Department of Energy%
{%
\setcounter{footnote}{0}%
\renewcommand{\thefootnote}{\fnsymbol{footnote}}%
\footnote{Contract number DE-AC02-98CH10886}
}
for providing facilities essential 
for completion of this work.


\begin{thebibliography}{*}

\bibitem{Teper}
For a review of glueball mass calculations, see:
M.J.~Teper,
hep-th/9812187.

\bibitem{PDG}
Particle Data Group, Eur. Phys. J. 3 (1998).

\bibitem{Lee}
See, for example:
W.~Lee and D.~Weingarten,
Nucl.\ Phys.\ Proc.\ Suppl.\ {63} (1998) 194.

\bibitem{Kirk}
See, for example: A.~Kirk
WA102 Collaboration,
hep-ph/9908253,
to be published in the proceedings of the International Europhysics
Conference
on High-Energy Physics (EPS-HEP 99), Tampere, Finland, 15-21 Jul 1999. 

\bibitem{Ochs}
See, for example:
W.~Ochs,
hep-ph/9909241,
to be published in the proceedings of the International Europhysics
Conference
on High-Energy Physics (EPS-HEP 99), Tampere, Finland, 15-21 Jul 1999.

\bibitem{JE}
{J. Ellis, Nucl. Phys. B22 (1970) 478;}\\
{R.J. Crewther, Phys. Lett. B33 (1970) 305.}

\bibitem{JE1}
{R.J. Crewther,  Phys. Rev. Lett. 28 (1972) 1421;\\
M.S. Chanowitz and J. Ellis, Phys. Lett. B40 (1972) 397; 
Phys. Rev. D7 (1973) 2490.}

\bibitem{sch}
{J. Schechter, Phys. Rev. D21 (1980) 3393;\\
A. Salomone, J. Schechter and T. Tudron, Phys. Rev. D23 (1981) 1143.} 

\bibitem{EL}
{J. Ellis and J. Lanik, Phys. Lett. B150 (1985) 289.}

\bibitem{NSVZ}
{V.A. Novikov, M.A. Shifman, A.I. Vainshtein and V.I. Zakharov, 
Nucl. Phys. B191 (1981) 301.}
\bibitem{scale}
{J. Collins, A. Duncan and S.D. Joglekar, Phys. Rev. D16 (1977) 438;\\
N.K. Nielsen, Nucl. Phys. B120 (1977) 212.}

\bibitem{FK}
{H. Fujii and D. Kharzeev, hep-ph/9903495; Phys. Rev. D, {\it in press}.}
\bibitem{EFK}
{J. Ellis, H. Fujii and D. Kharzeev, {\it in preparation}.}
\bibitem{Ioffe}
{B.L. Ioffe, Nucl. Phys. B188 (1981) 317 [Erratum: B 191 (1981) 591];\\
L.J. Reinders, H. Rubinstein and S. Yazaki, Phys. Rep. 127 (1985) 1. }
\bibitem{SVZ}
M.A. Shifman, A.I. Vainshtein and V.I. Zakharov, Nucl. Phys. B147 (1979) 385.

\end{thebibliography}
\end{document}